\documentclass[aps,prl,twocolumn,10pt, superscriptaddress]{revtex4-2}

\usepackage{bm}
\usepackage{epsfig}
\usepackage{bbold}
\usepackage{graphicx}
\usepackage{amsmath}
\usepackage{amssymb}
\usepackage{amsbsy}
\usepackage{color}
\usepackage{subfigure}
\usepackage{nicefrac}
\usepackage{slashed}
\usepackage{afterpage}
\usepackage{psfrag}
\usepackage{hyperref}

\usepackage[capitalise]{cleveref}
\usepackage{dsfont}
\usepackage{comment}
\newcommand{\beq}{\begin{equation}}
\newcommand{\eeq}{\end{equation}}
\newcommand{\bea}{\begin{eqnarray}}
\newcommand{\eea}{\end{eqnarray}}

\newcommand{\f}{\mathfrak{f}}

\allowdisplaybreaks

\begin{document}

\title{Gravitational waves from a dilaton-induced, first-order QCD phase transition}

\author{Aleksandr Chatrchyan}
\email{aleksandr.chatrchyan@su.se}
\affiliation{The Oskar Klein Centre for Cosmoparticle Physics,
Department of Physics,
Stockholm University, AlbaNova, 10691 Stockholm, Sweden
\looseness=-1}
\affiliation{Nordita, KTH Royal Institute of Technology and Stockholm University\\
Hannes Alfv\'ens v\"ag 12, SE-106 91 Stockholm, Sweden
\looseness=-1}
\author{M.C.~David~Marsh}
\email{david.marsh@fysik.su.se}
\affiliation{The Oskar Klein Centre for Cosmoparticle Physics,
Department of Physics,
Stockholm University, AlbaNova, 10691 Stockholm, Sweden
\looseness=-1}
\author{Charalampos Nikolis}
\email{charalampos.nikolis@fysik.su.se}
\affiliation{The Oskar Klein Centre for Cosmoparticle Physics,
Department of Physics,
Stockholm University, AlbaNova, 10691 Stockholm, Sweden
\looseness=-1}

\date{\today}

\begin{abstract}
\noindent
We show that 
a
`QCD dilaton' field, whose vacuum expectation value sets the strong coupling, can render the Quantum Chromodynamic (QCD) confinement transition first-order.
The QCD dilaton is cosmologically attracted to a false vacuum at weak coupling in the early universe. Quantum tunnelling towards the true vacuum triggers prompt chiral symmetry breaking and confinement of QCD, leading to detonating bubbles of the hadronic phase. We find that plasma sound waves produced by this dilaton-induced, first-order QCD phase transition generate a stochastic gravitational wave signal strikingly similar to the recently detected gravitational wave background from Pulsar Timing Arrays. We briefly comment on how this theory can be probed through collider experiments and cosmology.
\end{abstract}

\maketitle

Despite decades of enormous progress in observational and theoretical cosmology, the first microsecond of the Universe remains poorly known. Direct evidence for events taking place prior to Big Bang Nucleosynthesis (BBN) is scarce, which limits the ability to use cosmology as a `poor man's accelerator' \cite{Klauder:1972je, Linde:1990flp}.  The gateway to higher energies is provided by the Quantum Chromodynamic (QCD) phase transition (PT), in which the hot  Big Bang plasma transitioned from the high-temperature quark-gluon phase into the low-temperature confined, hadronic phase.  In the Standard Model (SM) of particle physics, the QCD PT is a smooth cross-over occurring at temperatures around $T_{\rm QCD}\sim160$ MeV \footnote{Given that the transition is a smooth cross-over, there is no sharply defined critical temperature, but proxies based on e.g.~the peak of the disconnected chiral susceptibility have been devised to define the cross-over temperature.}. Cross-overs are comparatively calm cosmic events, leaving few observational imprints. Thus, in the SM, the prospects for directly probing the QCD PT are dim.  

The QCD confinement PT is accompanied by chiral symmetry breaking and the growth of the strong coupling constant, $g_3$.   
In the Standard Model, $g_3$ is perturbatively small at high energies and runs to strong coupling at around the QCD scale, which through dimensional transmutation is given by
\beq
\Lambda_{\rm QCD} = \mu \, {\rm exp}\left[ -\frac{8\pi^2}{\beta_0 g^2_3(\mu)}\right] \, ,
\eeq 
at one-loop order, where $\mu$ is an energy scale, $\beta_0 = 11- \tfrac{2}{3}N_f$ for $N_f$ flavours.  
The ultraviolet value of the coupling constant,  $g_3^{\rm UV}= g_3(\mu_{\rm UV})$, is, like all Standard Model parameters, assumed to be constant: once fixed, it is unchanging over cosmic time. 

By contrast, quantum gravity is expected to have no free parameters or true constants. All parameters of low-energy effective theories, like the Standard Model, are expected to be secretly field-dependent. We refer to the scalar field, $\phi(x)$,  that makes the strong coupling constant dynamical as the `QCD dilaton’ (or just dilaton, for short). 

In this paper, we show that the dynamics of $\phi$ can effectively turn the QCD phase transition first-order, leading to detonating bubbles of the confined phase, separated from the quark-gluon plasma by a dilaton domain wall. The expanding bubbles quickly reach relativistic speeds and produce a stochastic gravitational wave background with an amplitude and spectrum that agree well with recent observations from Pulsar Timing Arrays (PTAs)~\cite{NANOGrav:2023gor, EPTA:2023fyk, Reardon:2023gzh, Xu:2023wog}, thus begging the question: \emph{Have we already observed the first signals from a dilaton-induced first-order QCD phase transition?}

The mechanism introduced in this Letter is based on multiple-field dynamics and quantum vacuum transitions, and should be contrasted to the much-studied scenario of \emph{thermal} PTs. It is well established that physics Beyond the Standard Model (BSM) can turn the QCD cross-over into a thermal first-order PT. If the resulting PT is sufficiently strong, such models can produce gravitational waves 
in the frequency band probed by PTAs \cite{Caprini_2010, Brandenburg:2021tmp,RoperPol:2022iel}. However, recent work has shown that the relevant parameter space is hard to realise in explicit models,
either due to weak transition strengths \cite{Witten:1984rs, Applegate:1985qt,agrawal2025supercooledconfinement}, non-relativistic wall velocities \cite{cline2025bubblewallvelocityfirstorder}, or failure to percolate \cite{Athron:2024}.
 Conversely, we will now show that the dilaton-induced QCD PT naturally leads to relativistic bubble walls and offers a compelling, and rather minimal, explanation of the PTA observations.

{\bf \emph{QCD and its dilaton}} --- The Lagrangian of QCD and its dilaton, $\phi$, is given by
\begin{align}
{\cal L}= -\frac{1}{2} \f\left(\phi\right) &{\rm tr}\left(G_{\mu \nu} G^{\mu \nu } \right) + \sum_i \bar q_i \left(i \slashed{D} - m_i\right) q_i \nonumber \\
&+\frac{1}{2} (\partial \phi)^2
- V_0(\phi) \, ,
\label{eq:dilaton-lagrangian}
\end{align}
where $G_{\mu \nu}= \partial_\mu A_\nu-\partial_\nu A_\mu - i[A_\mu, A_\nu]$, $q_i$ denotes quark species $i$, and $\slashed{D} = \gamma^\mu( \partial_{\mu} - i A_\mu)$ for the non-Abelian gauge potential,  $A_\mu$. The real-valued gauge kinetic function $\f\left(\phi\right)$ sets the QCD coupling constant as $\f\left(\langle \phi \rangle \right)= 1/(g^{\rm UV}_3)^2$ at some high energy scale.

We assume that the zero-temperature dilaton potential, $V_0(\phi)$, has two minima: the true vacuum at $\phi_{\rm TV}$ and the false vacuum at $\phi_{\rm FV}$, with $\Delta V = V_0(\phi_{\rm FV})- V_0(\phi_{\rm TV}) >0$, and,  importantly, $\Delta \f = \f(\phi_{\rm FV})- \f(\phi_{\rm TV}) >0$. This assumption means that QCD is more weakly coupled in the false vacuum than in the true vacuum. We do not model the microphysical origins of $V_0(\phi)$  and $\f(\phi)$, but note that the requisite structure can be readily realised with simple ingredients.  Scalar potentials with 
localised bumps (or dips) can, e.g., be generated through the Coleman-Weinberg potential by integrating out bosons (or fermions) whose masses are minimised at finite values of $\phi$.
In potential string theory realisations of our scenario,  $\phi$ would be a modulus field whose potential can be quite complex, receiving both perturbative and non-perturbative contributions \cite{mcallister2023modulistabilizationstringtheory,Ibanez:2012zz,Kaplunovsky_1995}. Stabilised string compactifications with positive vacuum energy automatically involve multiple local minima (some asymptotic in field space) \cite{Kachru:2003aw, Balasubramanian:2005zx}, and long-standing general arguments suggest that our universe is located in regions of field space where the scalar potential is rather complex, with multiple important contributions \cite{Dine:1985he, Denef:2008wq}, generating a rich landscape of local minima. Similarly, the gauge kinetic function may descend from the holomorphic gauge kinetic function of supergravity, which receives both one-loop and non-perturbative corrections \cite{Dine:1986vd, Weinberg:1998uv}. 
Below, we provide a concrete toy model 
that realises our scenario and draws on ingredients commonly found in string compactifications, however, many generalisations are possible. Recent work on related theories include~\cite{GarciaGarcia:2016xgv, vonHarling:2017yew, Ipek:2018lhm, Croon:2019ugf, Berger_2019, Bruggisser:2022rdm, Bruggisser:2022ofg,
Carenza:2024tmi, gao2024firstorderdeconfinementphasetransition, bhallaladd2025leptogenesiseraearlysu2}. 

In the present universe, $\phi=\phi_{\rm TV}$, and the properties of the dilaton are experimentally constrained.
Collider searches at the Large Hadron Collider (LHC) have been used to constrain dilaton theories with linear gauge kinetic functions, $f \sim \phi/\Lambda$, leading to limits on $\Lambda$ in the TeV-range for dilaton masses $2\, {\rm TeV} \leq m_\phi \leq  5\, {\rm TeV}$ \cite{Danielsson:2019ftq}.  In this work, we assume $m_{\phi \, {\rm TV}} \sim m_{\phi \, {\rm FV}} \sim 10\, {\rm TeV}$, for simplicity. Moreover, the QCD coupling constant is set by experiments to $\alpha_s(m^2_Z)=0.1180\pm0.0009$ (cf.~\cite{ParticleDataGroup:2024cfk} and references therein), which implies 
$$
\f_{\rm TV} \equiv \f\left(\phi_{\rm TV}\right)=\frac{\beta_0}{8\pi^2}\ln\left(\frac{\Lambda_{\rm UV}}{m_Z}\right)+\frac{1}{4\pi\alpha_s(m_Z)}\;.
$$
For example, a UV-scale of $\Lambda_{\rm UV}\sim 10^{19}$ GeV, implies $\f_{\rm TV}\simeq 4.2$ for $\beta_0=7$ (as appropriate for 6 quark flavours).

{\bf \emph{The dilaton-dependent QCD scale}} --- The weaker gauge coupling in the false vacuum implies an exponentially lower strong-coupling scale, $\Lambda_{\rm QCD}$, than in standard QCD: 
\beq
\label{eq:Lambda-diff}
\Lambda^{\rm FV}_{\rm QCD} = \Lambda^{\rm TV}_{\rm QCD} \, {\rm exp}\left(- \frac{8\pi^2 \Delta \f}{\beta_0}  \right) \, .
\eeq
For $\Delta \f \sim {\cal O}(1)$, QCD with $\phi=\phi_{\rm FV}$ remains weakly coupled well below the Standard Model cross-over temperature $T^{\rm TV}_{\rm QCD}$, suggesting that chiral symmetry breaking and confinement would only happen at $T^{\rm FV}_{\rm QCD}\ll T^{\rm TV}_{\rm QCD}$.

Below the confinement temperature (for any $\phi$), we model the low-energy QCD dynamics with a two-quark linear sigma model  featuring an explicit chiral symmetry breaking \cite{Kapusta:2006pm}
\beq
V_{\rm \chi}\left(\phi,\chi\right)=\frac{\lambda}{4}\left(\chi^4-2f_\pi^2\chi^2\right)-m^2_\pi f_\pi \chi\;,
\label{eq:VQCD}
\eeq
with $\chi$ the chiral condensate and $f_\pi(\phi),\; m_\pi(\phi)$ the pion decay constant and the pion mass, respectively. The quantities $f_\pi,\;m_\pi$ inherit their $\phi$-dependence from their dependence on the confinement scale, 
and obey the scaling relations (cf.~\cite{Croon:2019ugf} for a similar discussion)
\bea
\frac{f_\pi\left(\phi_{\rm FV}\right)}{f_\pi\left(\phi_{\rm TV}\right)}= 
\frac{m_\pi(\phi_{\rm FV})}{m_\pi(\phi_{\rm TV})}
=
\exp\left[-\frac{8\pi^2 \Delta\f}{\beta_0} \right]\;.
\eea
Note that, in \eqref{eq:VQCD}, we have omitted a $\chi$-independent (but $\phi$-dependent) contribution $\sim f_\pi^4$, as this contribution can be absorbed into $V_0(\phi)$.  Taking $\lambda=20$ reproduces the condensate mass of $m_\chi=600$ MeV \cite{Scavenius:2000bb}.

 More elaborate effective theories modelling the confinement transition  in addition to the chiral symmetry breaking can be constructed \cite{RevModPhys.64.649,Schaefer:2007pw}; however,  the simpler linear sigma model suffices to capture the relevant physics of our scenario, which we now discuss.

{\bf \emph{A dilaton-induced, first-order QCD phase transition}} --- Early in cosmic history, the dilaton was in  thermal equilibrium with the quark-gluon plasma and its dynamics was governed by the temperature-dependent effective potential
\beq
V_{\rm eff}(\phi)=V_0(\phi)+V_{\rm T}(\phi)+V_{\cal P}(\phi)\;,
\label{eq:Veff}
\eeq
where 
$$
V_{\rm T}(\phi)=\frac{T^4}{2\pi^2}J_{\rm B}\left(\frac{m_\phi^2(\phi)}{T^2}\right)\;
$$
denotes the one-loop thermal effective potential due to dilaton self-interactions \cite{quiros1999finitetemperaturefieldtheory}. Here, $J_B$ is the thermal bosonic function and $m_\phi(\phi)$ is the field-dependent mass. The term $V_{\cal P}(\phi)$ represents the thermal QCD pressure-corrections ($\mathcal{P}_{\rm QCD}=-V_{\cal P}$) in a fixed dilaton background  for $N_c=3$  and $N_f$ flavours  \cite{Laine_2016}:
\begin{align}
\nonumber V_{\cal P}(\phi)=&-\frac{8\pi^2}{45} T^4\Bigg(\left(1 + \frac{21 N_f}{32}  \right)
\\& -\frac{15\alpha_s(\phi,T)}{4\pi}\left( 
1 + \frac{5 N_f}{12}\right)+\mathcal{O}(\alpha_s^{3/2})\Bigg)\; \; ,
\label{eq:VP}
\end{align}
where $\alpha_s(\phi)=1/(4\pi \f(\phi))$. The contribution of the pressure term due to the coupling of the dilaton to the Standard Model in the context of PTs was discussed recently in \cite{knappperez2025thermalhistorynonequilibratedscalars} and in the context of dark matter in \cite{Cyncynates:2024ufu}.\footnote{ 
This correction is valid for the $\phi$ values we are interested in at temperatures higher than the confinement scale, since the RG-evolved coupling satisfies $\alpha_s(\phi_{\rm TV},\mu=T)<1$, so that QCD is still perturbative.  For temperatures where QCD is non-perturbative near $\phi_{\rm TV}$, we use the fit to lattice calculations from \cite{HotQCD:2014kol} (c.f.~the Supplemental Material \cite{SuppMat} for a more detailed discussion).}  \footnote{We note that the interactions of dilaton fluctuations with the gluons with vertex structure $\sim (p_1^\mu p_2^\nu-g^{\mu\nu})$ renormalise the couplings but do not have finite contributions at $T\neq0$ that could change the shape of the 1-loop thermal potential, and hence we neglect these contributions.}.

\begin{figure*}[t]
    \centering
    \includegraphics[width=0.47\textwidth]{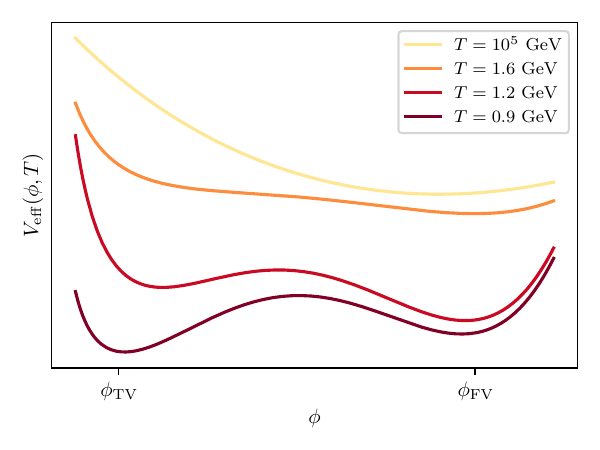}
    \hfill
    \includegraphics[width=0.47\textwidth]{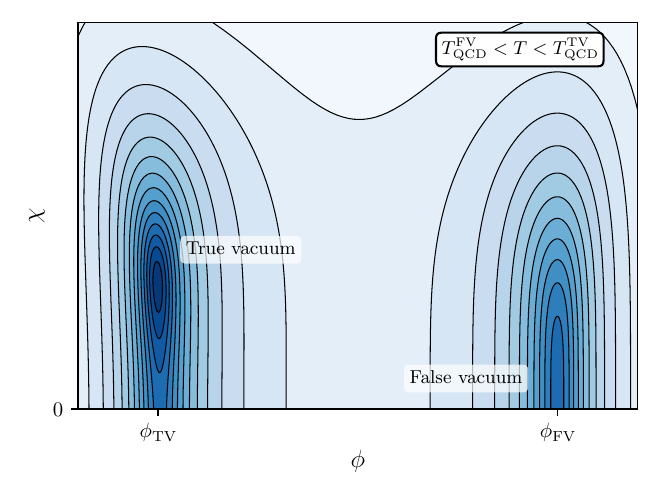}
    \caption{{\bf Left:} The dilaton potential, $V_{\rm eff}(\phi)$, at different temperatures. The values of the potential are rescaled for visual clarity. For $T\sim100~\mathrm{TeV}\gg m_\phi$ the minimum is located in between the tree-level vacua. As temperature drops ($T\sim \mathcal{O}({\color{blue}1})~\mathrm{GeV}\ll m_\phi$), the minimum  moves towards the $\phi_{\rm FV}$ while a local minimum emerges near $\phi_{\rm TV}$. When temperature corrections become unimportant, $\phi_{\rm TV}$ becomes the global minimum. {\bf Right:} 
    Schematic contour plot for the two-field potential, $V_{\rm tot}(\chi, \phi)$ for $T_{\rm QCD}^{\rm FV}<T<T_{\rm QCD}^{\rm TV}$.
    }
        \label{fig:thermal_evolution}
        \label{fig:Vphichi}
\end{figure*}

The effective potential of Eq.~\eqref{eq:Veff} is 
shown schematically in the left panel of Fig.~\ref{fig:thermal_evolution} for four different temperatures. At sufficiently high temperatures ($T\gg m_\phi$),  $V_{\rm eff}$ is dominated by the stabilising, leading-order contribution from $V_T$, 
$\sim m^2(\phi)T^2$, and the loop-corrected part of $V_{\cal P}$, which push the dilaton towards weak coupling (corresponding to larger field values).  
The potential has a single minimum  in the range $\phi_{\rm TV}<\phi <\phi_{\rm FV}$, slightly offset towards $\phi_{\rm FV}$.

At intermediate temperatures, the thermal contributions from self-interactions become Boltzmann-suppressed and can be ignored, and $V_{\rm eff}$ develops a second minimum at smaller field values. Notably, 
due to $V_{\cal P}$, the large-field minimum approaching $\phi_{\rm FV}$ is the global minimum of $V_{\rm eff}$ at these temperatures, and the 
dilaton is attracted to the false vacuum.  This is a key property of our scenario and a general feature of theories with field-dependent couplings: the theory is pushed to weaker couplings at high temperatures. This effect traps the dilaton at $\phi_{\rm FV}$, providing natural initial conditions for the weak-to-strong PT.

At sufficiently low temperatures, $V_{\rm eff}$ becomes dominated by the zero-temperature potential $V_0(\phi)$, and $\phi_{\rm TV}$ becomes the new global minimum of the potential. However, the dilaton remains classically trapped at $\phi_{\rm FV}$. The right panel of Fig.~\eqref{fig:Vphichi} shows a schematic representation of the full two-field potential, $V_{\rm tot}\left(\phi,\chi\right)=V_{\rm eff}(\phi)+V_{\rm \chi}(\phi,\chi)$,
for 
$
T_{\rm QCD}^{\rm FV}<T<T_{\rm QCD}^{\rm TV}
$.
 For $\langle\phi\rangle = \phi_{\rm TV}$, chiral symmetry is broken by 
 $\langle \chi \rangle \neq 0$ in the vacuum. 
 For $\langle \phi \rangle=\phi_{\rm FV}$, QCD is in the unconfined phase, and, consistently,  $\langle \chi \rangle=0$ and chiral symmetry is unbroken.

With $\phi$ in the false vacuum, the quark-gluon plasma may cool below $T_{\rm QCD}^{\rm TV}$ without triggering the cross-over transition to the hadron gas (which for  $\phi=\phi_{\rm FV}$ would happen around $T_{\rm QCD}^{\rm FV}$). However, the dilaton vacuum is non-perturbatively unstable and eventually decays to $\phi_{\rm TV}$ through quantum tunnelling, proceeding through the nucleation of true-vacuum bubbles. The nucleation rate per unit volume is given by 
$
   \Gamma/\mathcal{V}=A e^{-S_4}
$
\cite{Coleman:1977py,Callan:1977pt},
with $S_4$ the time independent $O(4)$ symmetric bounce action, and $A$ a prefactor. 
The nucleation temperature, $T_n$, is bounded from above by the requirement that the global minimum of $V_{\rm eff}$ is at $\phi_{\rm TV}$. This gives $T_n \ll m_\phi$ and, more stringently, $T_n < T_c$, where $T_c$ denotes the `critical temperature' at which the two minima have the same energy: $V_0(\phi_{\rm TV}) + V_{\cal P}(\phi_{\rm TV}, T_c) = V_0(\phi_{\rm FV}) + V_{\cal P}(\phi_{\rm FV}, T_c)$. Lattice simulations have found that the QCD pressure at strong coupling is lower than 
the corresponding ideal gas pressure \cite{Karsch_2000, HotQCD:2014kol} (see \cite{Petreczky_2012} for a review). Parametrically, we expect $T_c^3T_{\rm QCD} \sim \Delta V$ when ${\cal O}(300)\;{\rm MeV} \lesssim T_c \lesssim 2\; {\rm GeV}$ and  $T_c^4 \sim \Delta V$ for higher critical temperatures, cf.~the Supplemental Material \cite{SuppMat} for a more detailed discussion. 
Moreover, the nucleation temperature is bounded from below by the BBN temperature $T_n\gtrsim T_{\rm BBN} \approx 4\; {\rm MeV}$ \cite{Kawasaki:1999na, Kawasaki:2000en, Hasegawa:2019jsa, Barbieri:2025moq}. For $\Delta {\frak f} \sim {\cal O}(1)$, $T_{\rm BBN}\gg  T_{\rm QCD}^{\rm FV}$, so the false-vacuum QGP does not undergo a QCD cross-over transition prior to the dilaton tunnelling.  
In the main text of this Letter, we explore  $T_n$ in the range 
\begin{align}
 T_{\rm BBN} < T_n < T_{\rm QCD}^{\rm TV} \, .
 \label{eq:Thierarchy}
\end{align}
The complementary range, $T_{\rm QCD}^{\rm TV}<T_n < T_c$, naturally predicts ultrarelativistic `runaway' bubbles that, upon collision, generate gravitational wave signals at higher frequencies. Such models can be probed by future gravitational wave detectors, but do not affect the QCD cross-over transition, cf.~the Supplemental Material for a more detailed discussion \cite{SuppMat}.
Fixing the nucleation temperature corresponds to fixing the decay rate through the approximate equation 
$\Gamma/{\cal V}\simeq H^4(T_n)$,
expressing that, at $T_n$, one bubble is nucleated per Hubble volume and Hubble time. 

Outside the bubbles, the plasma remains in the pre-confined quark-gluon phase (as $T_n>T_{\rm QCD}^{\rm FV}$), but \emph{inside} the bubbles, the temperature is lower than the local QCD confinement scale (i.e.~$T_{\rm QCD}^{\rm TV}$), leading to prompt hadronisation and chiral symmetry breaking. In the two quark linear sigma model, the approximate timescale of confinement can be found from $t_{\rm roll} \sim {\cal O}(1/m_{\chi}) \ll H^{-1}$, with $m_{\chi}=\sqrt{2\lambda f_\pi}$, so chiral-symmetry breaking inside the true-dilaton-vacuum bubbles is approximately instantaneous. Thus, in this scenario, the QCD PT is no longer a smooth cross-over, but characterised by expanding bubbles of the hadronic phase inside the quark-gluon plasma; {\it{this scenario effectively turns the QCD phase transition first order.} }   We stress, however, that the mechanism is intrinsically quantum and  multiple-field, and lies outside the much-studied framework of thermal QCD PTs.

We require that the transition percolates (at plasma temperature $T_\star$) which, for a PT with constant $\Gamma$, requires $\alpha\lesssim 20$ \cite{Freese_2022}, where $\alpha=\Delta V/\rho_{\rm rad}(T_{\rm n})$. 
For such transitions, the inverse duration of the PT, $\beta$, is set by $\alpha$ and satisfies $8\gtrsim \beta/H_*>3$~\cite{Freese_2022}. 
As the energy density $\Delta V$ is channelled into the Standard Model, the plasma is reheated to the temperature
  $T_{\rm rh} = (1+ \alpha)^{1/4} T_n$,
 here assuming 
$T_\star \approx T_n$. We restrict our discussion to $T_{\rm rh} \lesssim T_{\rm QCD}^{\rm TV}$, so that the hadrons do not deconfine (and then undergo a `second' QCD  transition, this time a cross-over).

The subsequent cosmology and observational predictions of this scenario depend on the speed at which the bubbles expand. As we show in the Supplemental Material~\cite{SuppMat}, the bubble walls reach ultrarelativistic terminal velocities before  the bubbles collide.  This is in notable contrast to BSM or holographic models of thermal first-order QCD PT, in which the latent heat is typically too small to overpower the pressure from the quark-gluon plasma, thereby making ultrarelativistic velocities unattainable.

{\bf \emph{Gravitational wave signal}} --- As the relativistic dilaton bubbles grow large, collide, and percolate, they produce a stochastic gravitational wave (GW) background, potentially observable today~\cite{Witten:1984rs, hogan1986gravitational}. This is particularly interesting given the recent observation of the Hellings-Down correlation in Pulsar Timing Array (PTA) datasets~\cite{NANOGrav:2023gor, EPTA:2023fyk, Reardon:2023gzh, Xu:2023wog}, providing evidence for a stochastic GW background in the nanohertz range. Such a background could potentially be explained astrophysically through supermassive black hole binary mergers~\cite{NANOGrav:2020bcs}, but this explanation has difficulties with accommodating the observed amplitude and spectrum~\cite{middleton2021massive, izquierdo2022massive}.
By contrast, several studies have found that GWs generated by a first-order PT provides a better fit to the data~\cite{NANOGrav:2023hvm, EPTA:2023xxk, Bringmann:2023opz, Figueroa:2023zhu, Ellis:2023oxs, Addazi:2023jvg,Salvio_2023, fujikura2023nanogravsignaldarkconformal,PhysRevLett.132.221001, Salvio_2024,Winkler:2024olr, Fujikura_2024, Chen_2024, gonçalves2025supercooledphasetransitionsconformal, balan2025subgevdarkmatternanohertz} (see \cite{Brandenburg:2021tmp} for an alternative gravitational wave explanation generated by turbulent sources).

Since the bubbles are expected to reach a terminal velocity before they collide, the main contribution to the GW signal in this scenario is from the dynamics of sound waves generated in the plasma \cite{Hindmarsh:2013xza, Hindmarsh:2015qta,Caprini:2015zlo, Hindmarsh:2017gnf,Caprini:2018mtu} (the contribution to the signal coming from bubble wall collisions, which is relevant for runaway PTs, is discussed in the Supplemental Material~\cite{SuppMat}).
 
Estimates of the amplitude and spectrum of the GW background can be parametrised through rather simple formulas, depending on $T_\star\approx T_n,~\alpha,~\beta$, and $v_w$ (cf.~\cite{Caprini:2019egz} and, for a recent review, \cite{Hindmarsh:2020hop}).  
The peak frequency is given by  
\beq
    f_{\rm SW}=\frac{17.53~\text{nHz}}{v_w} \frac{\beta}{H_*}\frac{T_n}{100~\text{MeV}}\left(\frac{g_*}{61.75}\right)^{1/6}\;.
\eeq

Thus, the dilaton-induced first-order QCD PT proposed in this work results in peak frequencies in the ${\cal O}(10$--$100\, \text{nHz})$ range.

The corresponding GW energy spectra, $h^2\Omega_{\rm GW}$, give the fractional energy density per logarithmic frequency interval, rescaled by the squared Hubble parameter. 
The GW spectrum from sound waves is then 
\bea
\nonumber
\label{eq:GW-sw}
&h^2\Omega_{\rm SW}(f)=3.11
\times 10^{-6}\left(\frac{H_*}{\beta}\right)\left(\frac{\kappa_v \alpha}{1+\alpha}\right)^2\\& \times  \left(\frac{61.75}{g_*(T_\star)}\right)^{1/3}v_w\left(\frac{f}{f_{\rm sw}}\right)^3\left(\frac{7}{4+3\left(f/f{\rm sw}\right)^2}\right)^{7/2}\;,
\label{eq:GWsw}
\eea
where the efficiency factor  $\kappa_v$ has been inferred from simulations of ultra-relativistic walls \cite{Espinosa_2010},
$\kappa_v=\alpha/(0.73+\alpha+0.083\sqrt{\alpha})
$.
We note that Eq.~\eqref{eq:GWsw} is motivated by simulations of detonations with $\alpha \leq 0.67$ \cite{Correia:2025qif,caprini2025gravitationalwavesfirstorderphase}; for larger values of $\alpha$, it provides an often-used extrapolation.

\begin{figure}[t]
    \centering
    \includegraphics[width=0.47\textwidth]{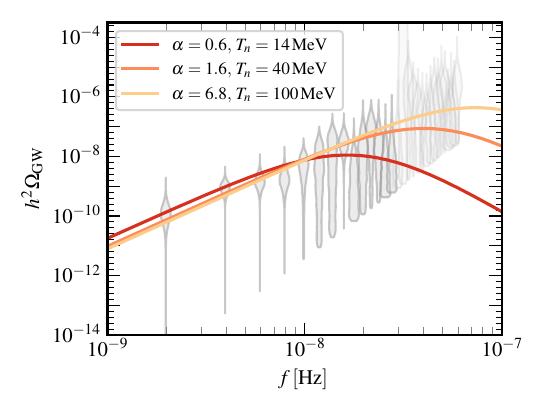}
    \caption{ Gravitational wave spectra from a dilaton-induced QCD PT overlaid with NANOGrav 15-year free-spectrum constraints (two shades of gray violins).  
    The red curve corresponds to the best-fit spectrum to the 14 lowest-frequency bins (darker gray). For more information, see the main text.
}
    \label{fig:NGtwopanel}
\end{figure}

In Fig.~\ref{fig:NGtwopanel} we show the GW energy spectrum generated by the dilaton-induced phase transition for several benchmark values of the parameters.  The spectra are overlaid with the posterior distributions from the NANOGrav 15-year free-spectrum analysis~\cite{NANOGrav:2023gor}, shown as violins in two shades of gray (see below).  
Remarkably, the predictions of our model are broadly compatible with the observed spectrum. In the Supplemental Material \cite{SuppMat}, we 
follow \cite{NANOGrav:2023hvm} and determine the parameter space favoured by the NANOGrav Data by performing a
Markov Chain Monte Carlo analysis using the PTArcade code~\cite{andrea_mitridate_2023, Mitridate:2023oar}.
Only the 14 lowest-frequency bins are used to constrain the spectrum, since the evidence for the GW background is dominated by the low-frequency modes \cite{NANOGrav:2023hvm}. These bins cover the frequency range $2\; {\rm nHz} \leq f \leq 28\; {\rm nHz}$ and are shaded in darker gray in Fig.~\ref{fig:NGtwopanel}. For clarity, we also plot the higher-frequency range up to $59\; {\rm nHz}$, shown in lighter gray, that is used for fitting the pulsar-intrinsic red noise.
The predicted spectrum provides an excellent fit to the data, with $68\%$ posterior credible intervals given by $11.7\; {\rm MeV}\leq T_n \leq 22.7\; {\rm MeV}$ and $0.47\leq \alpha \leq 0.84$. The best-fit spectrum is shown in red in Fig.~\ref{fig:NGtwopanel}, together with two illustrative examples of spectra for other parameter values. We note that the slight dip in the inferred spectrum above 20 nHz drives the preference for a low nucleation temperature. Higher values of $T_n$ push the peak of the spectrum to higher frequencies, while increasing $\alpha$ lowers the peak frequency and increases the overall strength of the signal.

{\bf \emph{An explicit model}} --- The dilaton-induced QCD PT can be realised in a range of models. Here, we consider one simple, explicit example, for concreteness. We defer the construction of an ultraviolet completion of this model to future work. We start from the Lagrangian 
\beq
\mathcal{L}\supset\Lambda_1^2\frac{\partial_\mu\varphi\partial^\mu\varphi}{2\varphi^2}-\frac{\varphi}{2\Lambda}\rm{tr}\, G_{\mu\nu}G^{\mu\nu}\;,
\eeq
for the energy scales $\Lambda_1$ and $\Lambda$. This general form of the Lagrangian, with non-canonical kinetic terms and a  linear dependence of the gauge kinetic function on $\varphi$, is often found in the four-dimensional supergravity description of string compactifications, where it is common for moduli fields to have logarithmic K\"ahler potentials and linear holomorphic gauge kinetic functions at tree level. 

In terms of the canonically normalised dilaton field, $\phi = \Lambda_1 \ln (\varphi/\Lambda)$, the Lagrangian is given by 
\beq
\mathcal{L}\supset \frac{1}{2}\partial_\mu\phi\partial^\mu\phi-
\frac{e^{\phi/\Lambda_1}}{2}\rm{tr}\, G_{\mu\nu}G^{\mu\nu}\;,
\eeq
i.e.,~with $\f= {\rm exp}(\phi/\Lambda_1)$. 

An exponential hierarchy between $\Lambda_{\rm QCD}^{\rm TV}$ and $\Lambda_{\rm QCD}^{\rm FV}$ is generated if $8\pi^2 \Delta \f /\beta_0 \gtrsim {\cal O}(1)$, which gives
$$
\frac{8\pi^2}{\beta_0} \Delta \phi f'_{\rm TV} = \frac{8\pi^2 \f_{\rm TV}}{\beta_0} \frac{\Delta \phi}{\Lambda_1} \gtrsim {\cal O}(1) \; ,
$$
where we have Taylor expanded $\f$ to leading order. For $\f_{\rm TV} = 4.2$ this gives a weak lower bound on the separation of the vacua: $\Delta \phi/\Lambda_1 \gtrsim 3 \cdot 10^{-2}$. We assume a toy-model effective potential for $\phi$ given by
\bea
&V_0(\phi)&=\frac{m^2}{2\Delta\phi^2}\left(\phi-\phi_{\rm TV}\right)^2\left(\phi_{\rm FV}-\phi\right)^2\\\nonumber &&+\frac{\epsilon(\phi-\phi_{\rm FV})}{\Delta \phi}
+ V_{\rm np}(\phi) 
\;,
\eea
for which we treat $\phi_{\rm TV}, \phi_{\rm FV}, m^2$ and $\epsilon$ as free parameters. 
Note that  $\Delta \phi = \phi_{\rm FV}-\phi_{\rm TV}$, and $m^2_{\rm TV/FV}=m^2$.
The term $V_{\rm np}$ includes contributions that become important for small field values, including the contribution $\sim f_\pi^4(\phi)$ from the linear sigma model, the gluon condensate $\sim -\frac{\beta_0}{32}\Lambda_{\rm QCD}^4(\phi)$ \cite{SHIFMAN1979385}, and further non-perturbative contributions of the form $\sim ce^{-8\pi^2\f(\phi)/\beta_0}$. The potential $V_{\rm np}$ is necessary to ensure that the minimum for the two-field potential $V_{\rm tot}(\phi,\chi)$ is located at $(\phi_{\rm TV},f^{\rm SM}_\pi)$. Finally, we denote as $~\Delta V\equiv\epsilon+V_{\rm np}(\phi_{\rm FV})-V_{\rm np}(\phi_{\rm TV})$.

To realise the hierarchy of Eq.~\eqref{eq:Thierarchy} with a nucleation temperature of $T_n=100\, {\rm MeV}$ one may, for example, choose $m \sim 10 \, {\rm TeV}$, $\Delta\phi\simeq2\times10^{-3}~{\rm GeV}$, $\Delta\phi/\Lambda_1\sim\mathcal{O}(1)$ and $\Delta V\simeq 0.7\rho_{\rm rad}(T=100\text{MeV})$.  In this case, the barrier height in $\phi$ direction is large, and the thin-wall approximation can be used to estimate $S_4$ and the decay rate.
Thus, our scenario can be realised with a rather minimal set of ingredients.

{\bf \emph{Discussion}} --- We have shown that quantum tunnelling of the QCD dilaton can turn the QCD phase transition first order and generate a gravitational wave signal intriguingly similar to that observed by Pulsar Timing Arrays. Notably, in our scenario, the PT is induced by multi-field dynamics rather than generated from thermal fluctuations, and its predictions differ quantitatively and qualitatively from the much-studied framework of thermal, first-order QCD transitions.

Our theory can be probed by collider experiments. Dilatons produced through gluon fusion can be searched for through dijet resonances. Data from Run 2 of the Large Hadron Collider (LHC) has already been used to constrain the theory up to $m_\phi \sim 5\, {\rm TeV}$ \cite{Danielsson:2019ftq}. Data from LHC Run 3, the planned High-Luminosity LHC, and future collider experiments at higher energy will probe some of our theory's most interesting parameter space. 
 
A first-order QCD PT may source inhomogeneities in the plasma at the onset of Big Bang Nucleosynthesis (BBN), which can affect the abundances of heavy elements \cite{PhysRevD.31.3037, Applegate:1987hm} (see also \cite{bagherian2025bearableinhomogeneitybaryonasymmetry}). We expect to explore the impact of our dilaton-induced QCD PT on BBN in future work. 

An interesting extension of our theory includes the axion, which can be regarded as the pseudo-scalar partner of the dilaton. The mass of the axion increases exponentially from the false vacuum to the true, and an ambient `seed' abundance of axions can be trapped in relic pockets of the false vacuum state \cite{Carenza:2024tmi} (see also \cite{Witten:1984rs, Hindmarsh:1991ay}). If $\Delta \f$ is sufficiently large, such `axion relic pockets' can be stable and may inherit a substantial fraction of the energy density, $\Delta V$, leading to an overclosure of the universe unless $\alpha \ll 1$. For moderate $\Delta \f$, the axion relic pockets are transient `droplets', and provide an novel source of small-scale isocurvature.  

A further extension of this theory includes an axion-like particle (ALP)
coupled to a dark sector confining gauge theory in addition to QCD (i.e.~a `QCD ALP' \cite{Lella:2024dmx}). Such theories often suffer from a problem where stable cosmic domain walls come to dominate the energy density, in conflict with observations. Drawing on the findings of \cite{Blasi_2022,Blasi:2023sej}, we note that this problem may be solved if the combined ALP and dilaton tunnelling `eats' the domain walls, which may be achieved in theories where $\f(\phi)$ is non-monotonic.  We expect to make this mechanism more explicit in future work.

 While the explicit model presented in this work illustrates how it may be realised with rather simple ingredients, it would be interesting to embed this toy model in a more complete theoretical framework.

Finally, the mechanism proposed in this Letter may find applications outside cosmology. Emergent gauge theories with tunable `dilaton' parameters have been discovered in several condensed matter systems \cite{PhysRevLett.127.117205, doi:10.1126/sciadv.aav7444, Pikulin_2016, YU2021195}, and may potentially host non-trivial PTs similar to our scenario. 
 More generally, the essence of our proposal is that the order-parameter space of certain theories can naturally be augmented, leading to rich multiple-field dynamics that can transform the nature of predicted PTs.

\bigskip

{\bf{\emph{Acknowledgments}}}---We thank Mark Hindmarsh, Filippo Sala, and Javier Subils for stimulating discussions. 
This work was supported by the Swedish Research Council (VR) under grants 2018-03641, 2019-02337 and 2024-04289. This article is based upon work from COST Action COSMIC WISPers CA21106, supported by COST (European Cooperation in Science and Technology).

\bibliography{sample}
\appendix
\clearpage
\onecolumngrid
\begin{center}
\textbf{\large Supplemental Material}
\end{center}
\title{Supplemental Material for "Gravitational waves from a dilaton-induced, first-order QCD phase transition"}

\setcounter{section}{0}
\renewcommand{\thesection}{S\arabic{section}}

\setcounter{figure}{0}
\renewcommand{\thefigure}{S\arabic{figure}}

\section{Bubble wall velocity}

In this section, we calculate the pressure exerted on the wall by the quark-gluon plasma and determine the terminal speed of the bubble walls.  
Recall that we consider percolating phase transitions with time-independent $\Gamma$. The $\beta$-parameter is given by $\beta=\dot{I}(t_{\star})$, with $I(t)$ the average number of bubbles in the past comoving light cone \cite{Freese_2022}, i.e.~
\beq
\beta=\frac{4\pi\Gamma}{a(t_\star)}\int_0^{t_\star}dt' a^3(t') r^2_{\rm com}(t',t_\star)\;,
\eeq
where $a(t)$ denotes the scale factor, $r_{\rm com}(t_\star,t)$ the comoving bubble radius and $t_\star$ the percolation time, defined by $I(t_\star)=1$. As mentioned in the main text, for transitions that percolate, $0<\alpha\lesssim 20$, which corresponds to $8\gtrsim \beta/H_{\star}>3$

The bubble wall velocity is determined by the balance of the driving pressure, $\mathcal{P}_{\rm dr}\equiv \Delta V$,  and the pressure caused by friction from the surrounding plasma, $\mathcal{P}_{\rm fr}$. The calculation of bubble wall velocities in confining, first-order PTs is an active area of research \cite{Schwaller_2015, Ahmadvand_2018, Baldes:2020kam, Bigazzi_2021, PhysRevD.104.L121903,PhysRevD.104.L121903, PhysRevD.110.023509, Morgante_2023,han2023stochasticgravitationalwavesproduced, He_2024, Shao_2025, zheng2025quantitativeanalysisgravitationalwave,cline2025bubblewallvelocityfirstorder, agrawal2025supercooledconfinement}. In this work, the driving pressure is generated by the dilaton potential instead of the confinement dynamics. The expanding bubbles are in the ballistic limit, 
meaning that the
particles in the plasma do not interact significantly over the width of the wall, which in the plasma frame is given by $L_w/\gamma_w \sim 1/(m_\phi \gamma_w)$ where  $\gamma_w$ the Lorentz factor of the wall. The ballistic limit holds when $L_w/\gamma_w  \ll L_{\rm MFP}$, 
where $L_{\rm MFP}$ is the mean-free path of the particles in the quark-gluon plasma outside the bubbles  \cite{Ai_2025}. Parametrically, we expect $L^{-1}_{\rm MFP}\sim \alpha^2_{\rm s}(\phi_{\rm FV})\;T_n$ from $2\rightarrow2$ scatterings. Since QCD remains perturbative in the false vacuum, the ballistic approximation holds for all values of $\gamma_w$.

The dilaton bubble walls quickly accelerate to ultra-relativistic velocities. Indeed, 
assuming that quarks and gluons gain masses $\Delta m_q \sim \Delta m_g \sim\Lambda_{\rm QCD}$ when passing from the deconfined into the confined phase, the ballistic `one-to-one' pressure ($\mathcal{P}_{1\to1}$) from particles passing through or being reflected off the wall was calculated in  \cite{Ai_2025} and 
reduces to the form of \cite{Bodeker_2009} in the ultrarelativistic limit,
\beq
\mathcal{P}_{1\to1} = \frac{T_n^2}{48}\left( g_q^* \Delta m_q^2 +  2 g_g^* \Delta m_g^2   \right) < \alpha \rho_{\rm rad} =
\mathcal{P}_{\rm dr} \; , 
\eeq
where $g_q^*,\, g^*_g$ denote, respectively, the effective degrees of freedom for quarks and gluons in the plasma.  The inequality assumes $\alpha \gtrsim 3\cdot 10^{-3}\,  (\Lambda_{\rm QCD}/T_n)^2$, which holds across the most relevant parameter space.
Thus, one-to-one processes alone do not suffice to stop the runaway behaviour \cite{Bodeker_2009} \footnote{
A more accurate description of friction in a confinement phase transition is given in \cite{Gouttenoire:2021kjv}, and results in the scaling
$
\mathcal{P}_{\rm friction}=\mathcal{P}_{\rm BS}\sim\sum_{i\in BS}\frac{m_i^2T_n^2}{24}
$, 
which coincides with that of~\cite{Bodeker_2009}, used here.  }.
 
Gluon radiation is an additional source of friction.  Here, we model the pressure from gluon emission at the wall interface by the equivalent expression for  $W$ and $Z$ boson emission when entering bubbles of the low-temperature phase in models where the electroweak PT is first order~\cite{Bodeker_2017} (we explore an alternative scenario with runaway walls in the next section).  
Adopting the results of \cite{Bodeker_2017, Caprini:2019egz}, the gluon emission  pressure is given by  
\beq
\mathcal{P}_{1\rightarrow2}\simeq N_g\Delta m_g\gamma \;\alpha_s(\phi_w)T_n^3\;,
\eeq
with $\phi_w$ a value of the dilaton on the wall and $N_g=8$ is the number of gluons. Here, we have assumed that QCD remains perturbative at the wall, i.e.~for an intermediate value of $\phi$. This is typically justified since 
$\Lambda_{\rm QCD}(\phi_w) \ll \Lambda^{\rm TV}_{\rm QCD}$. The pressure contribution $\mathcal{P}_{1\rightarrow2}$ grows with $\gamma$ and eventually catches up with $\Delta V$, leading to an ultra-relativistic terminal velocity with gamma factor \footnote{Another potential friction source is bound state formation from string fragmentation~\cite{Baldes:2020kam}. However, due to insignificant supercooling, we expect confinement to happen among quark pairs rather than quarks forming flux-tubes attached to the bubble wall, and string fragmentation to be unimportant. Even in this scenario though, the leading and next-to-leading order friction from gluons has a similar dependence on $\gamma$}
\beq
\gamma_t=\frac{\Delta V-\mathcal{P}_{1\rightarrow 1}}{\alpha_s(\phi_w)N_g\Delta m_g T_n^3}
\approx
85\, \alpha \left(\frac{0.03}{\alpha_s(\phi_w)} \right)
\left(\frac{T_n}{\Lambda_{\rm QCD}}\right)
\; . 
\eeq
By comparing the size of the bubbles when reaching the terminal velocity to their separation $\beta^{-1}$, it is clear that the terminal velocity is reached well before bubbles collide.

\section{Gravitational waves from runaway walls}
In the main text, we have discussed the GW spectra in the well-motivated scenario where the dilaton bubble walls reach a terminal velocity before colliding. It is instructive to explore the alternate scenario in which the bubble walls `run away'. Such a scenario can be realised in models  that yield sufficiently small QCD coupling at the bubble wall interface along with large values of $\beta/H_*$ so that  bubbles never reach their terminal velocity before collision. When this happens, the dominant contribution to the gravitational wave spectrum is given by bubble wall collisions. As discussed in the main text, quantum phase transitions with $\Gamma\simeq$ constant yield $\beta/H_*\lesssim8$, which is not very large. However, if the dilaton phase transition is for some reason  driven by thermal fluctuations, larger values of $\beta/H_*$ can be attained, triggering runaway behaviour.

\begin{figure*}[t]
 
   \includegraphics[width=0.47\textwidth]{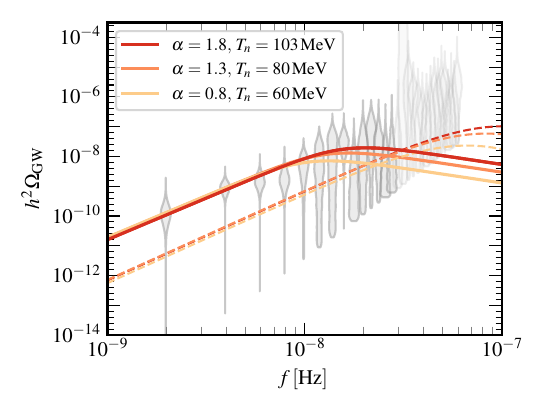}
    \caption{Gravitational wave spectra from a dilaton-induced QCD PT overlaid with NANOGrav 15-year free-spectrum constraints (gray violins), including the contribution from bubble wall collisions. Solid lines show the spectra in the regime where the energy is carried by bubble walls, while dashed lines correspond to the regime where the energy is converted into sound waves, for comparison. 
   The solid red curve corresponds to the best fit.}
    \label{fig:NGtwopanel_bc}
\end{figure*}

The contribution to the GW signal from bubble wall collisions is dominant if the walls do not convert much of their energy into the surrounding plasma. The GW spectrum in this case can be calculated using the envelope approximation~\cite{Kosowsky:1992rz, Kosowsky:1992vn, Huber:2008hg}, where it is assumed that only the uncollided regions of the expanding bubble walls contribute to the energy-momentum tensor~\cite{Caprini:2015zlo, Caprini:2018mtu},
\beq
\label{eq:GW-env}
\nonumber
h^2\Omega_{\rm coll.}(f)=1.96
\times 10^{-5}\left(\frac{H_*}{\beta}\right)^2\left(\frac{\kappa_{\rm coll.} \alpha}{1+\alpha}\right)^2 \left(\frac{61.75}{g_*(T_\star)}\right)^{1/3}\left(\frac{0.11 v_w^3}{0.42+v_w^2}\right)\left( \frac{3.8 \left(f/f_{\rm coll.}\right)^{2.8}}{1+2.8\left(f/f_{\rm coll.}\right)^{3.8}} \right)\;,
\eeq
with peak frequency given by \cite{Caprini:2015zlo}
\beq
\label{eq:coll-peak-freq}
f_{\rm coll}=\left( \frac{9.44~\text{nHz}}{1.8 - 0.1 v_w + v_w^2} \right) \frac{\beta}{H_*}\frac{T_n}{100~\text{MeV}}\left(\frac{g_*}{61.75}\right)^{1/6}\;.
\eeq
Here $v_w=1$ and $\kappa_{\rm coll}$ is the energy fraction that is converted into bubble walls. The corresponding spectra are shown in Fig.~\ref{fig:NGtwopanel_bc}, where we use several benchmark values of the parameters. 

The spectra are overlaid with the posterior distributions from the NANOGrav 15-year free-spectrum analysis~\cite{NANOGrav:2023gor}, shown as gray violins. For solid lines, we assume that all of the energy is in the form of colliding bubble walls, i.e.~$\kappa_{\rm coll }=1$. The dashed lines in Fig.~\ref{fig:NGtwopanel_bc} show the corresponding spectra arising from sound waves, which were discussed in the main text.

\section{The critical temperature}
\begin{figure*}[t]
    \centering
    \includegraphics[width=0.57\textwidth]{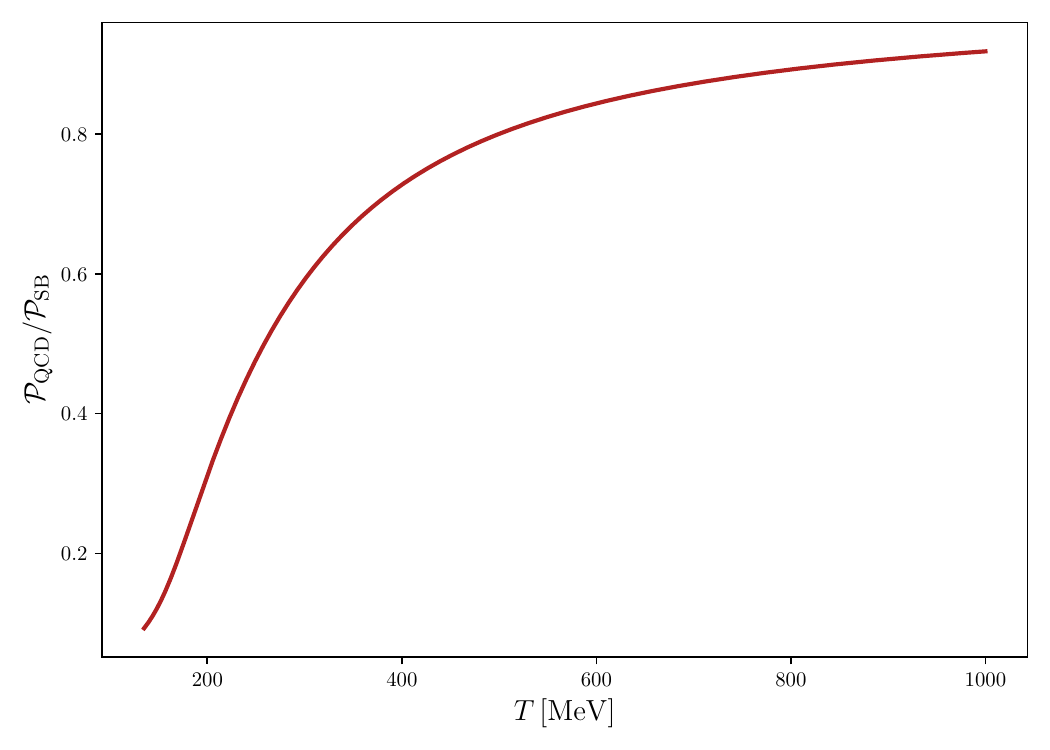}
    \caption{The fit of the QCD pressure as provided by the HotQCD collaboration \cite{HotQCD:2014kol}, for $T>T_{\rm QCD}=154$ MeV, normalised over the ideal gas pressure $\mathcal{P}_{\rm SB}=p_{\rm id}T^4$.} 
    \label{fig:PQCD}
\end{figure*}

In this section, we derive analytic estimates for the critical temperature, $T_c$, at which the two vacua (near $\phi_{\rm TV}$ and $\phi_{\rm FV}$, respectively) are degenerate. As discussed in the main text, this occurs at $T_c\ll m_\phi$ so that $V_{\rm eff}(\phi)=V_0(\phi)+V_{\cal P}(\phi)$. At sufficiently high temperatures relative to the QCD transition temperature,   $V_{\cal P}$ is given by the (negative of the) relevant ideal gas pressure, plus perturbative corrections, (cf.~Eq.~\ref{eq:VP}). At lower temperatures, non-perturbative lattice calculations are necessary to determine the pressure. We use the approximate semi-analytic formula from simulations of  (2+1)-flavour QCD by the HotQCD collaboration \cite{HotQCD:2014kol}, 
\begin{align}
{\cal P}_{\rm QCD} =
\frac{T^4}{2}\Big(1 + \tanh\left(c_t(t-t_0) \right) \Big)\times \frac{p_{\rm id} + a_n/t + b_n/t^2 + d_n/t^4}{1 + a_d/t+ b_d/t^2 + d_d/t^4} \, ,  
\label{eq:PhotQCD}
\end{align}
where $t= T/T_{\rm QCD}$, $T_{\rm QCD}= 154\; {\rm MeV}$, $p_{\rm id} = 95\pi^2/180$, $a_n= -8.77$, $a_d=-1.26$, and the remaining coefficients are given in Table II of \cite{HotQCD:2014kol}.
This expression, which is plotted in Fig.~\ref{fig:PQCD},  provides an excellent agreement with the data in the range $100 < T/{\rm MeV} < 400$, and a fair and, for our purposes sufficient, extrapolation with ${\cal O}(10\%)$ accuracy up to $T \sim 2\; {\rm GeV}$ \cite{HotQCD2018}. To leading order in $1/t$, which is relevant for the example discussed in the main text, the pressure is approximated by ${\cal P}_{\rm QCD}/T^4 \approx p_{\rm id} + (a_n -a_d p_{\rm id}) \tfrac{T_{\rm QCD}}{T}$.
With $\phi$ in the false vacuum, the theory is weakly coupled and the QCD pressure is given by the ideal gas pressure, as can be seen by replacing $T_{\rm QCD}$ by $T_{\rm QCD}^{\rm FV} \ll T_c$ in eq.~\eqref{eq:PhotQCD}. The pressure difference for $154\; {\rm MeV} \ll T \lesssim 2\; {\rm GeV}$ is then approximately given by
$V_{\cal P}(\phi_{\rm TV}, T) - V_{\cal P}(\phi_{\rm FV}, T) = (a_dp_{\rm id}- a_n) T_{\rm QCD} T^3$. Critical temperatures in this range are then given by
\begin{align}
    T_c = \left(\frac{\Delta V}{(a_dp_{\rm id}-a_n) T_{\rm QCD}}\right)^{1/3} \, .
    \label{eq:Tclow}
\end{align}
For lower critical temperatures,
the QCD pressure decreases rapidly relative to the ideal gas pressure, and 
$T_c$ can be found numerically by using the full expression of \eqref{eq:PhotQCD}. However, we note that for low $T_c$, the nucleation temperature often falls below the lower bound set by $T_{\rm BBN}$.   

By increasing $\Delta V$, one can achieve larger critical temperatures. For $T_c \gtrsim 2\; {\rm GeV}$, we use the perturbative, leading-order expression for the pressure difference: $V_{\cal P}(\phi_{\rm TV}, T) - V_{\cal P}(\phi_{\rm FV}, T) = \tfrac{8\pi^2}{45} T^4 \frac{15\alpha_s(\phi_{\rm TV}, T)}{4 \pi}\left(1 + \tfrac{5}{12} N_f \right)$, where we have used $\alpha_s(\phi_{\rm TV}, T)- \alpha_s(\phi_{\rm FV}, T) \approx \alpha_s(\phi_{\rm TV}, T)$. The critical temperature is then given by
\beq
T_c = \left( \frac{\Delta V}{\kappa \alpha_s(\phi_{\rm TV})}\right)^{1/4} \, ,
\label{eq:Tchigh}
\eeq
where $\kappa= \tfrac{2\pi}{3}  \left(1 + \tfrac{5}{12} N_f \right)$.

Equations \eqref{eq:Tclow} and \eqref{eq:Tchigh} provide upper bounds on the nucleation temperature, as $T_n < T_c$. In the main text we have discussed the case of $T_n < T_{\rm QCD}$. We now turn to the complementary case when $T_n>T_{\rm QCD}$.

\section{High-temperature transitions: $T_n > T_{\rm QCD}$}   
In this section, we show that dilaton phase transitions at high temperatures lead to runaway bubble walls that source gravitational wave signals that can be within the reach of space-based detectors. 

As $T_c \ll m_\phi$ in our scenario, the thermal population of $\phi$-quanta is Boltzmann suppressed by the time tunnelling becomes energetically permissible. As a result, even at $T_n \gg T_{\rm QCD}$, thermal transitions can be neglected and the transition remains quantum. For $T_n> T_{\rm QCD}$, QCD is deconfined on both sides of the bubble wall. It follows that gluons remain massless throughout the phase transition, and quark masses only change by a small amount across the walls, $\Delta m_q\propto \Delta \alpha_s\ll1$.   The dominant pressure on the bubble walls then comes from $1\rightarrow1$ processes, which, as mentioned in the main text, are insufficient to stop the runaway of the bubble walls. The resulting ultra-relativistic bubble walls can act as a `cosmic collider' \cite{shakya2024cosmiccollidershighenergy}, or may produce a forward shock front of ultra-relativistic gluons and quarks, potentially providing a concrete example for realising the `bubbletron' scenario of dark matter production \cite{Baldes:2023cih}.  We note that the subsequent QCD phase transition in this scenario will be a standard cross-over at $T_{\rm QCD}$.  

The rapidly expanding bubbles can source gravitational waves in a broad frequency range.  The peak frequency of the gravitational wave spectrum is given by eq.~\eqref{eq:coll-peak-freq}. For $v_w=1$, $300\, {\rm GeV}\lesssim T_n \lesssim 10^7\, {\rm GeV}$, and $3\leq \beta/H_\star \leq 8$, the spectrum peaks in the frequency range of upcoming space-based interferometers ($10^{-4} \leq f\; {\rm [Hz]}\leq 1$), such as LISA, TianQin and Taiji. Clearly, realising high nucleation temperatures requires a large dilaton mass, as $T_n \ll m_\phi$.

\section{Parameter space analysis}
To quantitatively determine how well our scenario compares to PTA observations, we employed a Markov Chain Monte Carlo analysis using the PTArcade code~\cite{andrea_mitridate_2023, Mitridate:2023oar} in the Ceffyl mode~\cite{lamb2023need} to derive posterior distributions for $\alpha$ and $T_n$ given the NANOGrav 15-year data.

Priors were chosen to span the ranges
$
10^{-1.5}< \alpha < 10^1, \quad 10^{-1.5}< \frac{T_n}{100 \rm MeV} < 10^1,
$
uniform in $\log$-space. The dilaton bubble walls reach ultrarelativistic speeds before collision, motivating $v_{w}=1$. We recall that $\beta/H_{\star}$ is determined uniquely from $\alpha$ for a time-independent nucleation rate. As discussed in the main text, only the first $14$ frequency bins were included in the analysis. 

\begin{figure*}[t]
    \centering
    \includegraphics[width=0.57\textwidth]{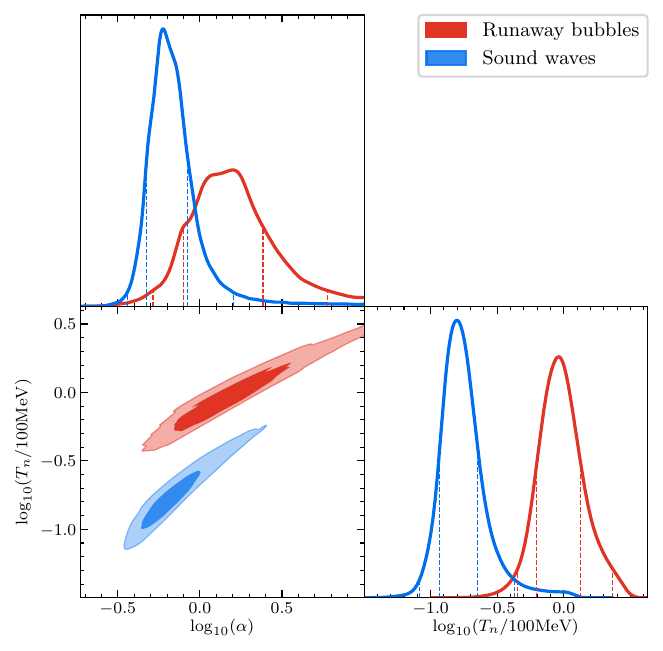}
    \caption{The diagonal panels show the marginalised 1D distributions, while the off-diagonal panel displays the 2D contours at $1\sigma$ and $2\sigma$  credibility levels for sound wave (blue) and runaway wall (red) scenarios. Raw data for this figure is available at \cite{dataset}.}
    \label{fig:posteriors}
\end{figure*}

Figure~\ref{fig:posteriors} shows the resulting posterior distributions, including the 2d $1\sigma$ and $2\sigma$ credible regions for gravitational waves generated by sound waves (blue) and bubble wall collisions (red). The best-fit parameters are $\alpha \approx 0.6$, $T_n\approx 14\, \rm MeV$ for sound waves and $\alpha \approx 1.8$, $T_n\approx\, 103\, \rm MeV$ for bubble wall collisions. 
 These parameters were identified as those corresponding to the maximum likelihood within the converged chain of $10^7$ samples.
The best-fit spectra provide excellent fits to the data. Following \cite{Winkler:2024olr}, we approximate the posterior distributions as piece-wise Gaussian for each bin, with different variance for upward and downward fluctuations, and re-calculate the best-fit parameters (finding small shifts to the values quoted above) and the goodness-of-fit. In this approximation,  we find $\chi^2/\text{degrees of freedom} =0.52$ and $0.55$, respectively, for the cases of sound waves and bubble wall collisions. 
 The number of degrees of freedom is $12 = 14-2$, where $14$ is the number of frequency bins and $2$ is the number of model parameters.
These values are comparable to those determined in \cite{Winkler:2024olr}, and show a consistent improvement over the astrophysical black hole binary merger explanation.      

\end{document}